\documentclass[prl,twocolumn]{revtex4}
\usepackage{graphicx}
\begin{document}
\unitlength=1mm
\title{Determining the Spectral Signature of Spatial Coherent Structures}
\author{L.R. Pastur, F. Lusseyran, Y. Fraigneau, B. Podvin}
\affiliation{LIMSI, University of Paris XI, 91403 Orsay Cedex,
France}
\date{\today}
\begin{abstract}
We applied to an open flow a proper orthogonal decomposition (pod)
technique, on 2D snapshots of the instantaneous velocity field, to
reveal the spatial coherent structures responsible of the
self-sustained oscillations observed in the spectral distribution
of time series. We applied the technique to 2D planes out of 3D
direct numerical simulations on an open cavity flow. The process
can easily be implemented on usual personal computers, and might
bring deep insights on the relation between spatial events and
temporal signature in (both numerical or experimental) open flows.

\vglue 0.3 truecm

\end{abstract}

\maketitle

One of the most challenging questions arising in open flows such
as jets, mixing layers, etc, is to understand the occurrence and
nature of robust and reproducible self-sustained oscillations
revealed in spatially localized time series, usually velocity or
pressure measurements. How such frequencies appear, and whether or
not they might be the signature of particular coherent spatial
patterns, still remain largely unresolved, although abundantly
documented \cite{flow,Huerre}. Such an understanding may moreover
appear of the upmost importance in control applications, in that
knowing which spatial event is generating such spectral signature
may lead to best fitted control scheme with respect to the
required goal. An example is given by flows over open cavities,
like in high speed trains, that generate very powerful
self-sustained oscillations that appear to be the main source of
noise emitted by the train. In that case, control will be aimed to
reduce or even suppress the source of noise, without reducing the
aerodynamic performances, and at the lowest energetic cost.

In this paper we \textit{(i)} show in a test case the ability of
the pod technique to associate self-sustained oscillations to
well-identified spatial coherent structures; \textit{(ii)}
confirm, as a consequence, the mixing layer origin of the most
energetic self-sustained oscillations in an open cavity flow. We
will show that 2D cuts out of the fully 3D flow are sufficient to
extract significant space-time events out of the flow. We are
using for that purpose a technique based on an empirical
decomposition of the flow, that optimizes a basis of (orthogonal)
eigen-modes with respect to the kinetic energy. The technique is
often known as to the proper orthogonal decomposition
(\textit{pod} hereafter) in the framework of fluid dynamics
\cite{Lumley}; or as the Karhunen-Lo\`eve decomposition in the
framework of signal processing \cite{KL}. (Other denominations
exist, such as empirical orthogonal decomposition, singular value
decomposition, etc, depending on the field of application
considered). To illustrate our point, we applied the technique to
3D direct numerical simulations of an air flow over an open cavity
\cite{cavity}.
The system is a cavity of length $L=10~cm$ along $x$ (the
longitudinal direction along which air is flowing), of depth $h=5$
cm (the aspect ratio $L/h$ is 2), and transverse size $l=20$ cm.
The cavity is enclosed into a vein $12$ cm high. The flow rate
velocity is $U_{0}=1.2$ m/s (Reynold's number Re $\simeq 8500$).
Simulations were performed following a finite volume approach
under an incompressible flow hypothesis. Spatial and time
discretization have a second order precision. The pressure field
is given by a Poisson's equation that requires a projection step,
such as to be in agreement with a non divergent velocity field. In
order to reduce the CPU time cost, the spanwise boundary
conditions are periodic. The $256\times 128\times 128$
mesh-spatial grid is refined in areas featuring strong velocity
gradients (boundary and shear layers)
--- with a mesh varying from 0.7 and $10~mm$ along the
longitudinal $x$ and vertical $y$ directions, and constant with
about $1.56$ mm over the transverse direction $z$ \cite{simul}.

Here we briefly expose the pod technique we implemented. The goal
is to compute the eigenmodes $\{\phi _n(t),\vec{\psi
}_n(\vec{r})\}$ that best fit the coherent structures composing
the flow, computed from a data base of $M$ different snapshots of
the velocity field, in such a way that any instantaneous snapshot
of the data base can be reconstructed by performing the sum over
the eigenmode basis:
\vspace{-4mm}
\begin{equation}
\vec{u}(\vec{r},t)=\sum _{n=1}^{M}\mu _n\phi _n(t)\vec{\psi
}_n(\vec{r}),\label{eq:decomposition}
\end{equation}
where the $\lambda _n=\mu _n^2$ are the eigenvalues of the
decomposition \cite{Lumley}. Typically $M$ was of the order of 600
frames.
Note that $\vec{u}$ being a
vector field, $\vec{\psi }$ must also be so; however we will also
use the notation $\psi $ when dealing with one component of the
field (usually it will be the longitudinal component along $x$).
A coherent structure can now be defined as an eigenmode of a
(2-pointwise linear) correlation matrix built on the data base
snapshots. There exists mainly two ways of building up a
correlation matrix: either performing a time correlation, or a
space correlation. With snapshots $\vec{u}(\vec{r},t)$ of size
$N=N_x\times N_y$ pixels (where $N_{x}\simeq 125$ and $N_y\simeq
100$ are respectively the snapshot dimensions along $x$ and $y$),
the space-correlation matrix
$$K(\vec{r},\vec{r}\ ')=\int _0^{t_M}u_p(\vec{r},t)u_q(\vec{r}\ ',t)dt$$
is of size $2N^2$ ($u_{p,q}$ are velocity components). We
restricted our analysis to the $x$,$y$-components of the velocity
field so as to mimic what is available from 2D experimental PIV
snapshots. On the contrary, the time correlation matrix
$$C(t,t')=\int \int
_\mathcal{S}\vec{u}(\vec{r},t)\vec{u}(\vec{r},t')d\vec{r}$$
is of size $M^2$ (if $M$ is the number of instant under
consideration), much smaller than $(2N)^2$ ($3.6\times 10^4$
against $4\times 10^8$). Keeping in mind that no more information
can be extracted from that contained in the data base itself, and
that at most $M$ relevant eigenmodes are therefore available from
the data set, we chose the second way (based on $C(t,t')$), known
as the \textit{snapshot pod} technique in the literature
\cite{legal,sirovich}. Practically, we start with a data base of
$M$ instantaneous spatial snapshots of the velocity field; in
experiments they can for example be obtained using PIV techniques
\cite{piv}. Then, the data are reshaped into a ``data matrix'' $A$
whose column elements are the pixels of a given snapshot. For that
purpose, each 2D snapshot is reshaped into a column vector (of
length $N$), by stacking over each other all the columns of the
snapshot, from the first to the last. Both $x$ and $y$ components
of the (vector) velocity field are further stacked in the same
column following the same procedure, starting with component $x$
at the top of the column, and then the component $y$ down to the
bottom of the column. The vertical size of $A$ is therefore $2N$.
The matrix $A$ contains as many columns as snapshots in the data
base (so that its horizontal dimension is $M$), the snapshots
being ranked from the left to the right of $A$ as the time is
flowing down. The matrix $A$ is therefore of dimension $M\times
2N$. The correlation matrix $C$ is next obtained by performing the
product $C=A^t\cdot A$, where $A^t$ is the transposed matrix of
$A$, and $\cdot $ the usual matrix dot product. (Note that the
space correlation matrix $K$ is given by $K=A\cdot A^t$) Applying
a singular value decomposition procedure on $C$, we obtain the
eigen-modes $\phi _n(t)$, rearranged as columns of a
\textit{chronos} \cite{legal} matrix $\Phi $ from left with $n=1$
to right with $n=M$. The spatial eigenmodes $\vec{\psi
}_n(\vec{r})$ (sometimes called \textit{topos} in the literature
\cite{legal}) are given following Eq.(\ref{eq:decomposition}) by
$\vec{\psi }_n(\vec{r})=\frac{1}{\mu _n}\int \phi
_n(t)\vec{u}(\vec{r},t)dt$. The $\vec{\psi }_n$ are reshaped into
columns of a \textit{topos} matrix $\Psi =(A\cdot \Phi )\cdot
D^{-1/2}$, following the same procedure as $A$, where $D$ is the
diagonal matrix of the eigenvalues $\lambda _n$, ranked from the
largest to the smallest value. The \texttt{Matlab}$^\copyright$
software is dedicated to matrix operations, so that the whole
process of building $A$, calculating $C$, performing the singular
decomposition to obtain the $\phi _n$, and determining the $\psi
_n$, takes, for $M=600$ and $N\simeq 37300$ no more than $30\ sec$
on a usual PC.

\begin{figure}[tbl]
\includegraphics[width=35mm]{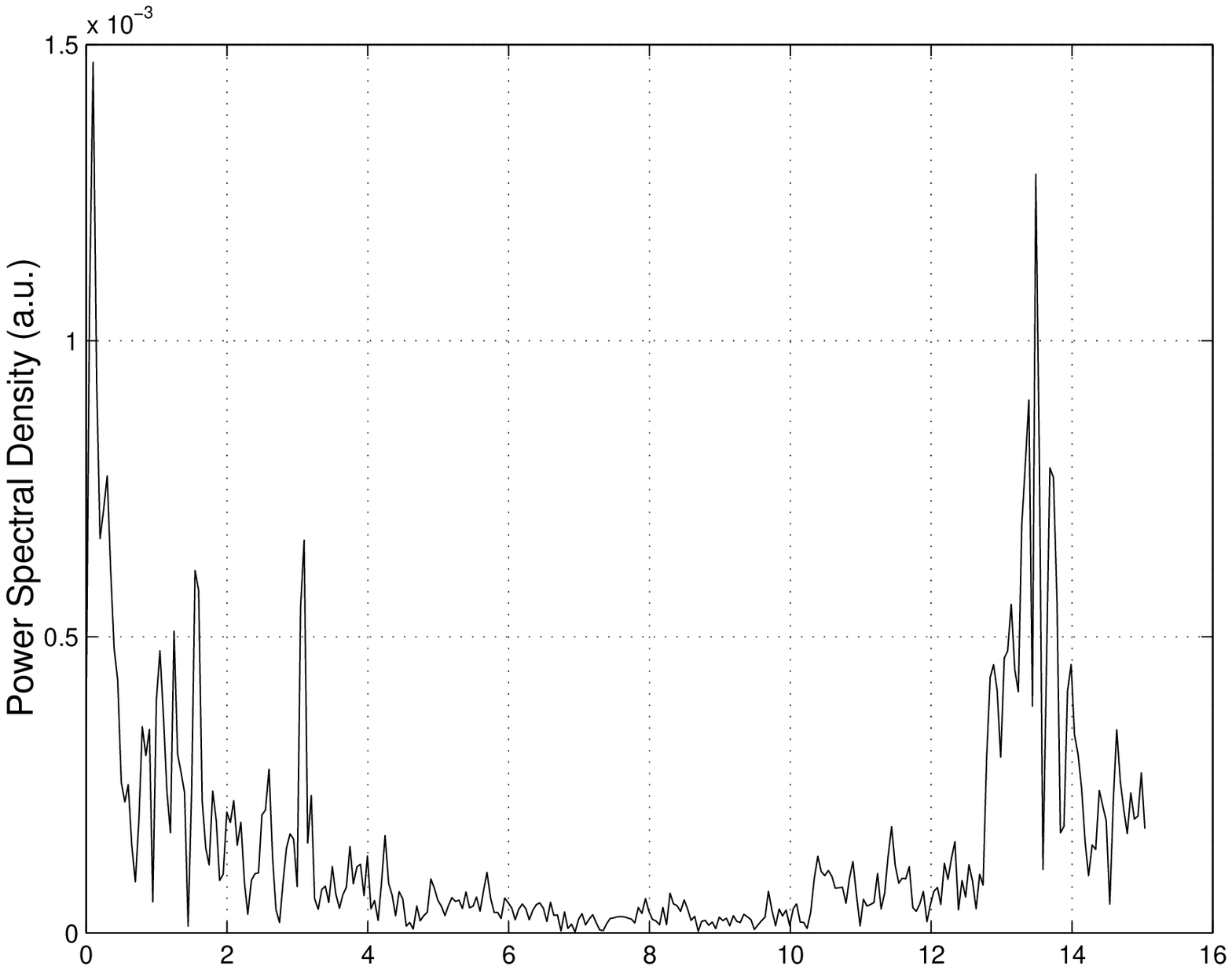}
\hfill
\includegraphics[width=35mm]{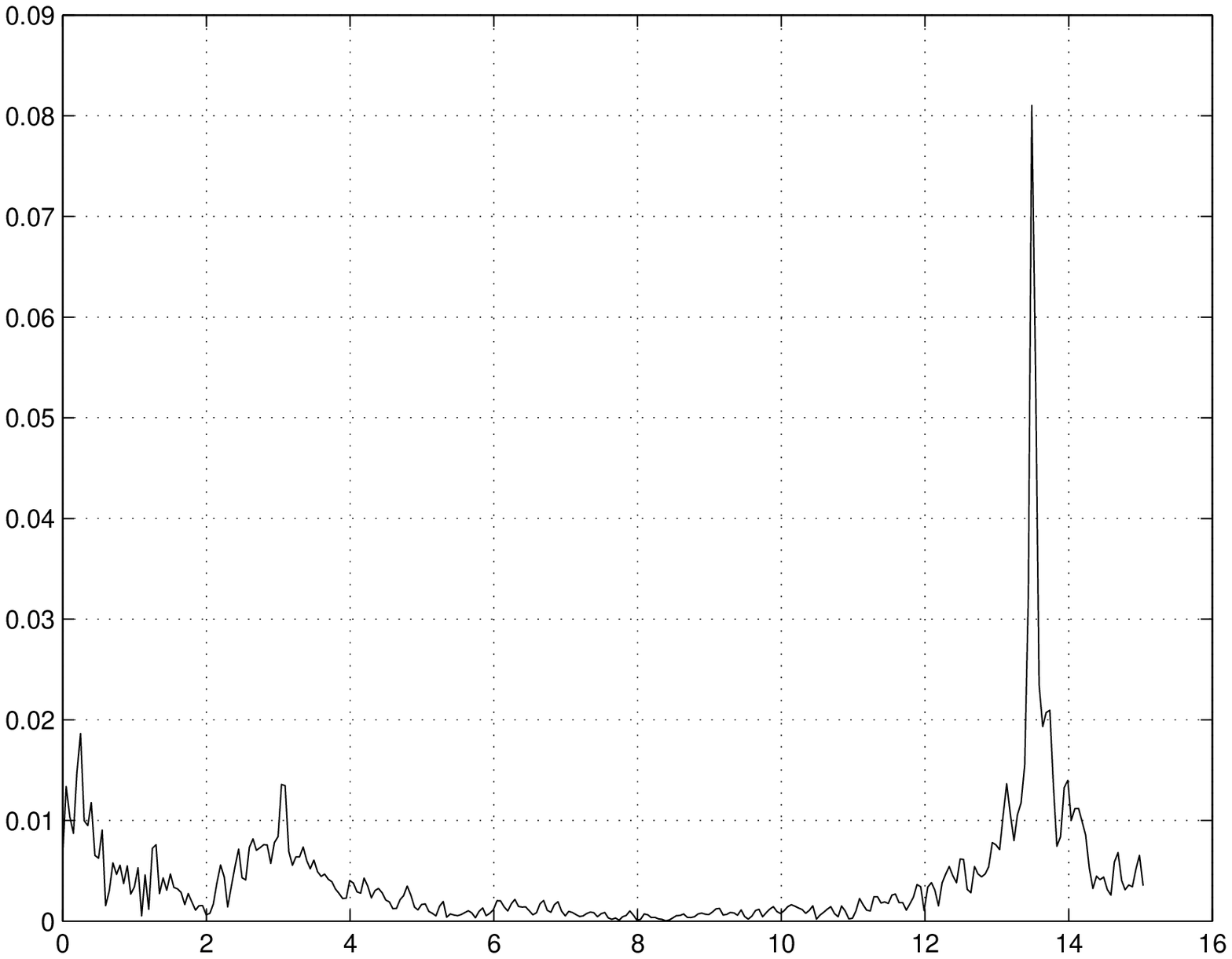}
\newline
\includegraphics[width=35mm]{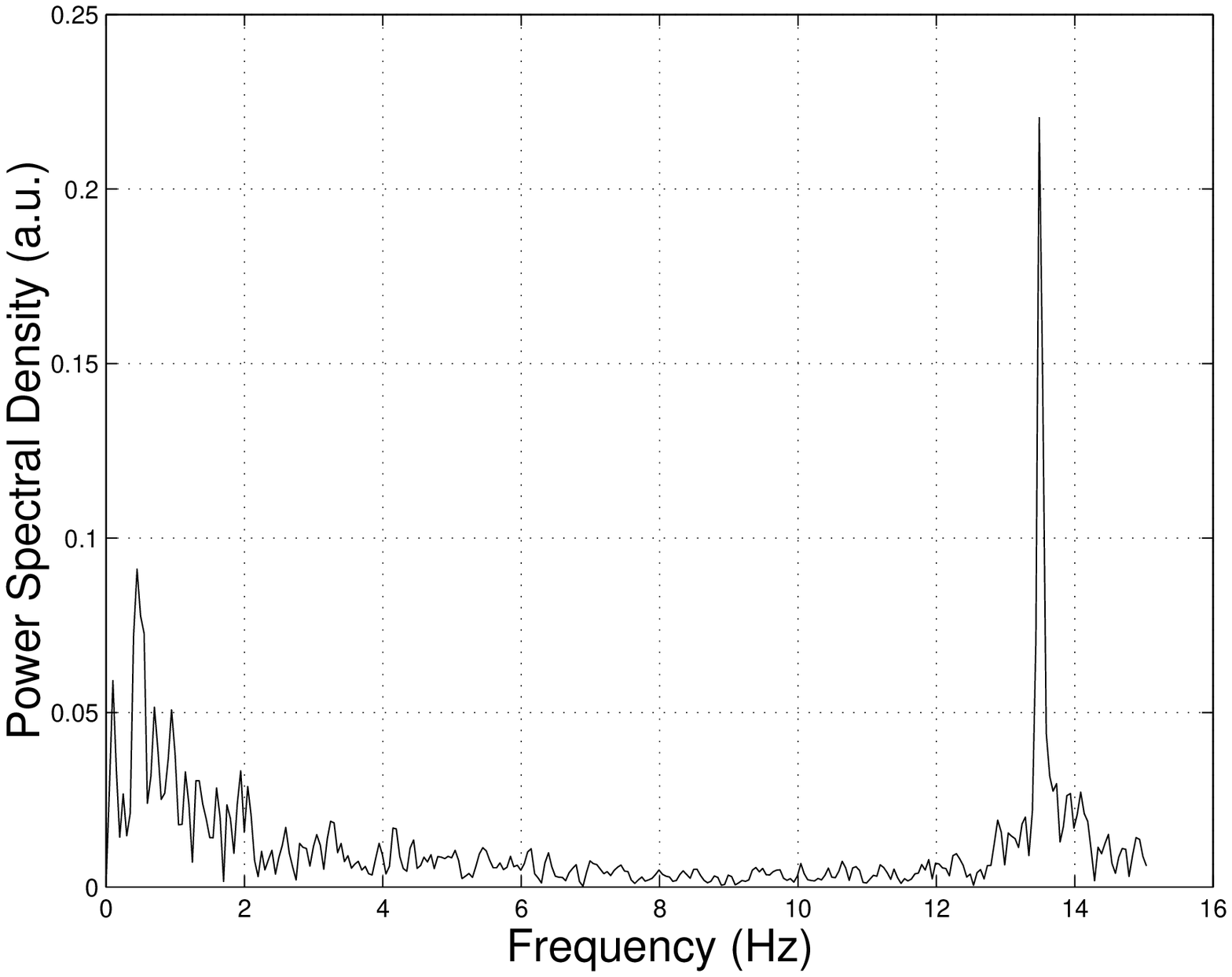}
\hfill
\includegraphics[width=35mm]{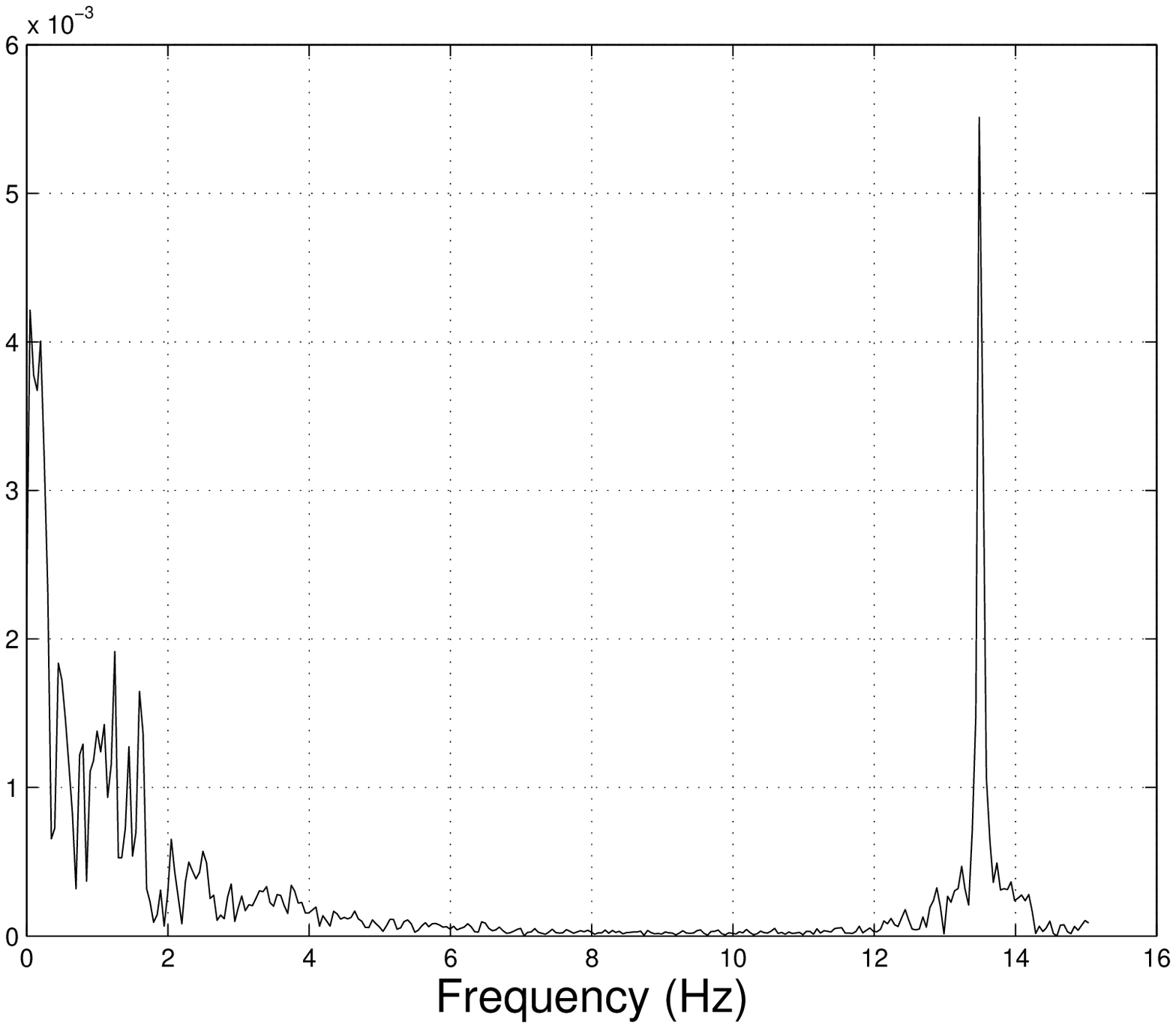}
\caption{Power spectral distribution of the $x$-component velocity
time series collected in the mixing layer, upstream and downstream
(top left to right), within the cavity, upstream and downstream
(bottom left to right), from 3D direct numerical
simulations.}\label{fig:tseries}
\end{figure}

We first present in Fig.\ref{fig:tseries} the spectral
distribution of time series provided by local recordings of one
component of the velocity field (here the longitudinal component
$u_x(t)$). Velocity recordings are done at 4 different locations:
2 within the mixing layer (one upstream, one downstream), and 2
within the cavity (upstream and downstream). In each of them
clearly appear peaks at about $f_0=13.5$ Hz (Strouhal number
St=1.06 when based on the cavity length $L$ and the reference
velocity; St=0.033 when based on the mixing layer thickness and
the mean velocity --- to be compared with the natural Strouhal
number St$_n$=0.03 of an unforced mixing layer \cite{Huerre}), and
it is now well accepted that this frequency is produced by the
instability of the mixing layer \cite{flow}. The spectral
component is recovered anywhere in the cavity, presumably due to
the overall pressure field coupling due to the fluid
incompressibility (the Mach number is about $4\times 10^{-3}$).

\begin{figure}[rtb]
\includegraphics[width=40mm]{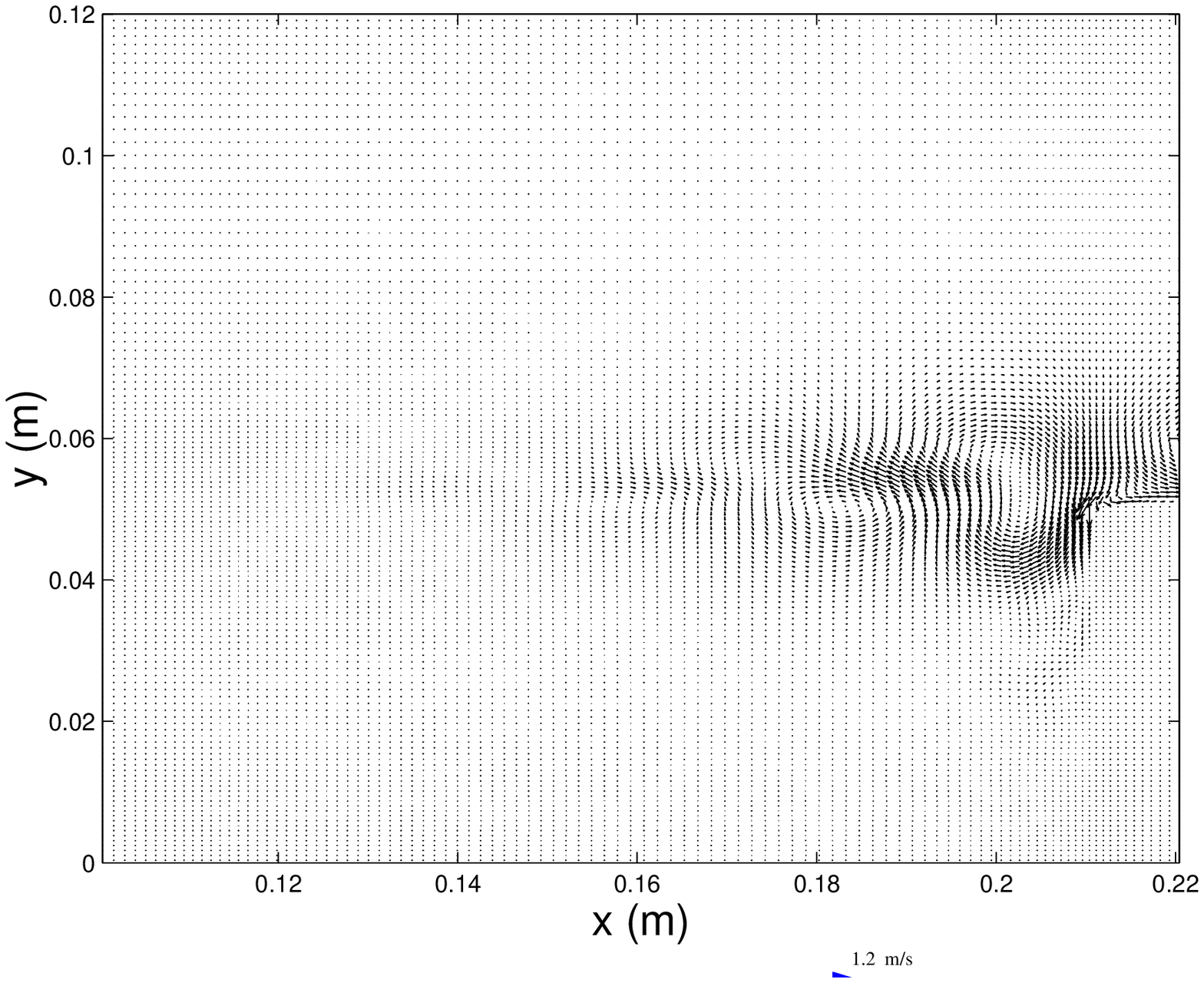}
\hfill
\includegraphics[width=40mm]{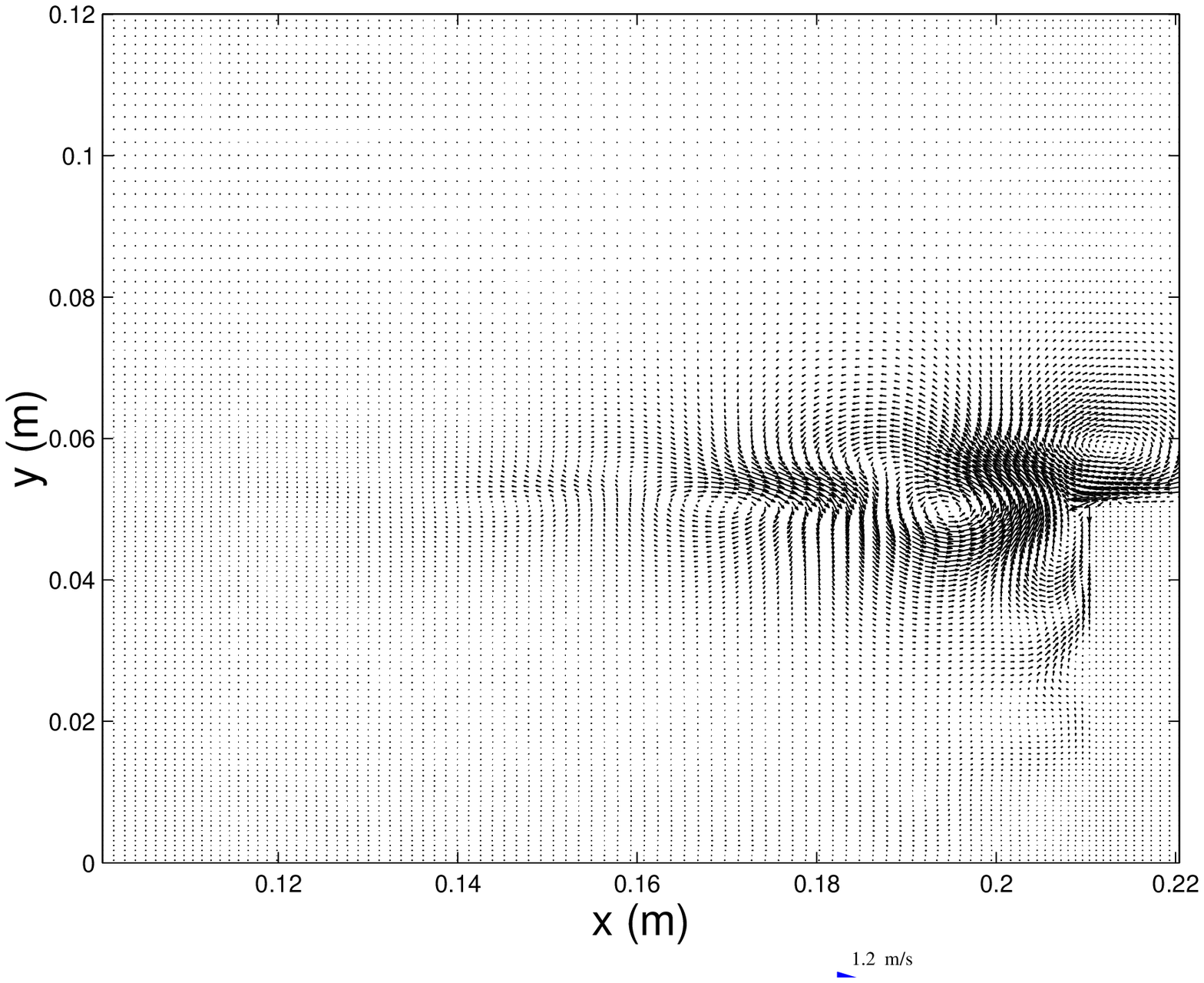}
\newline
\includegraphics[width=40mm]{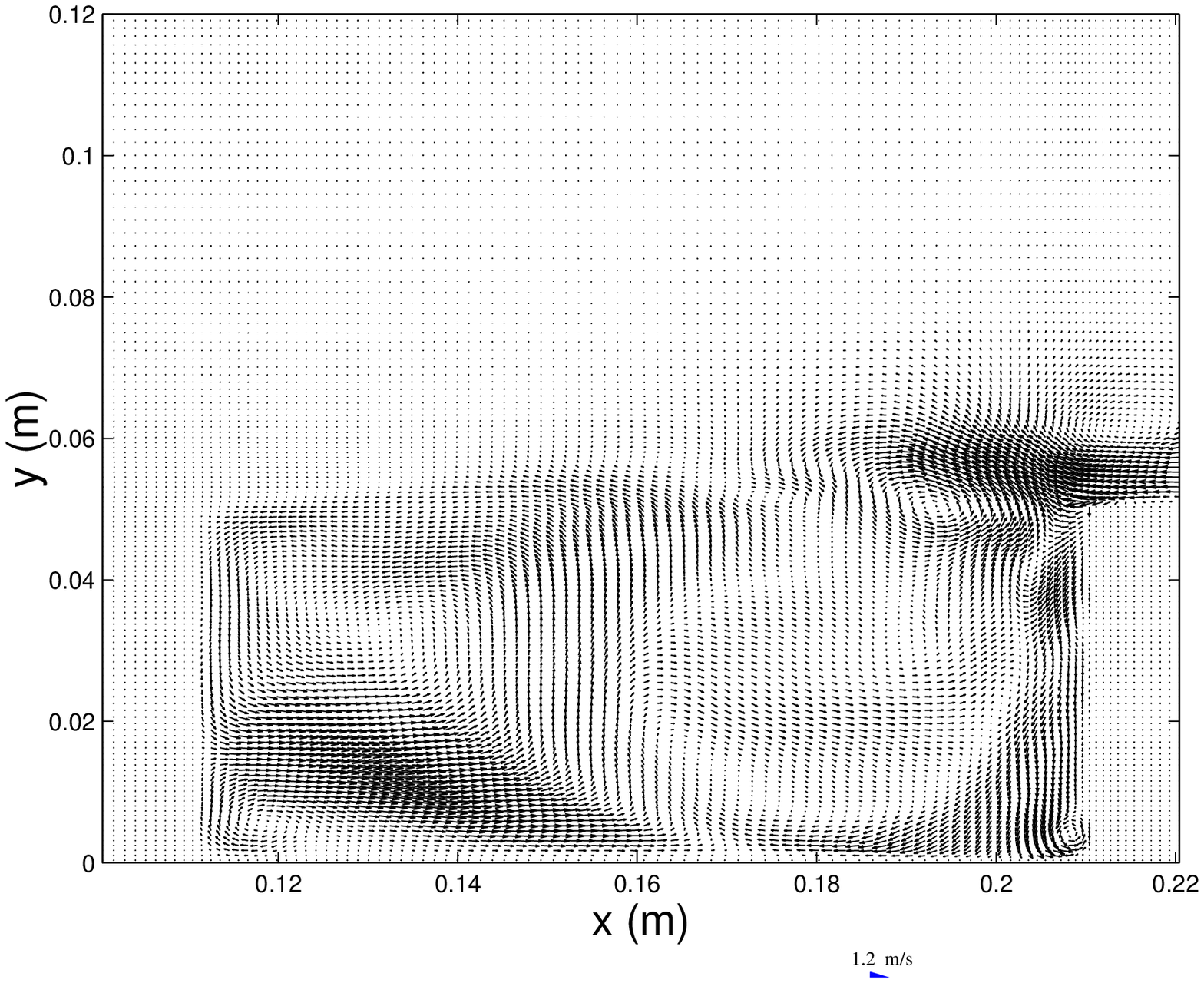}
\hfill
\includegraphics[width=40mm]{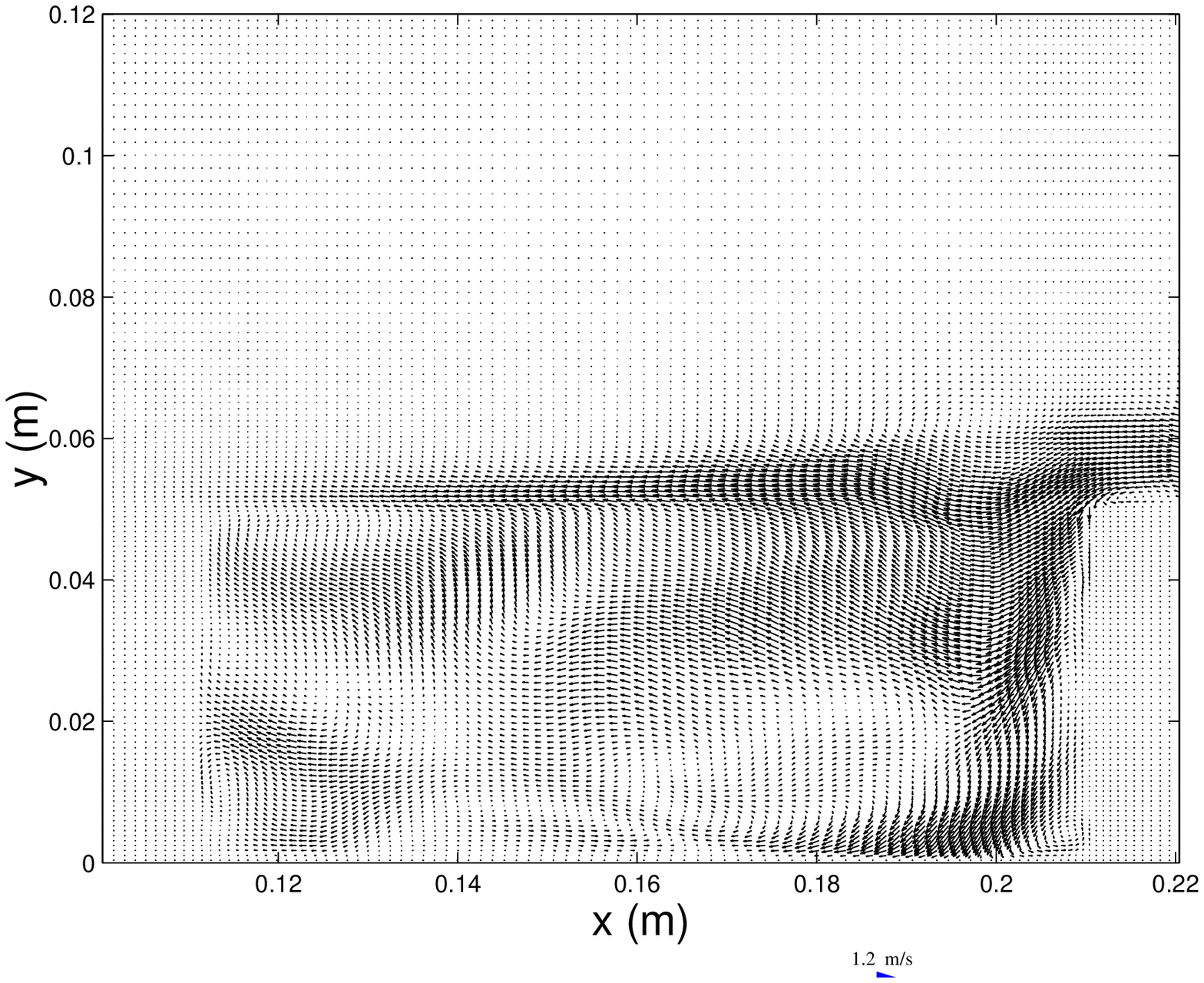}
\newline
\includegraphics[width=40mm]{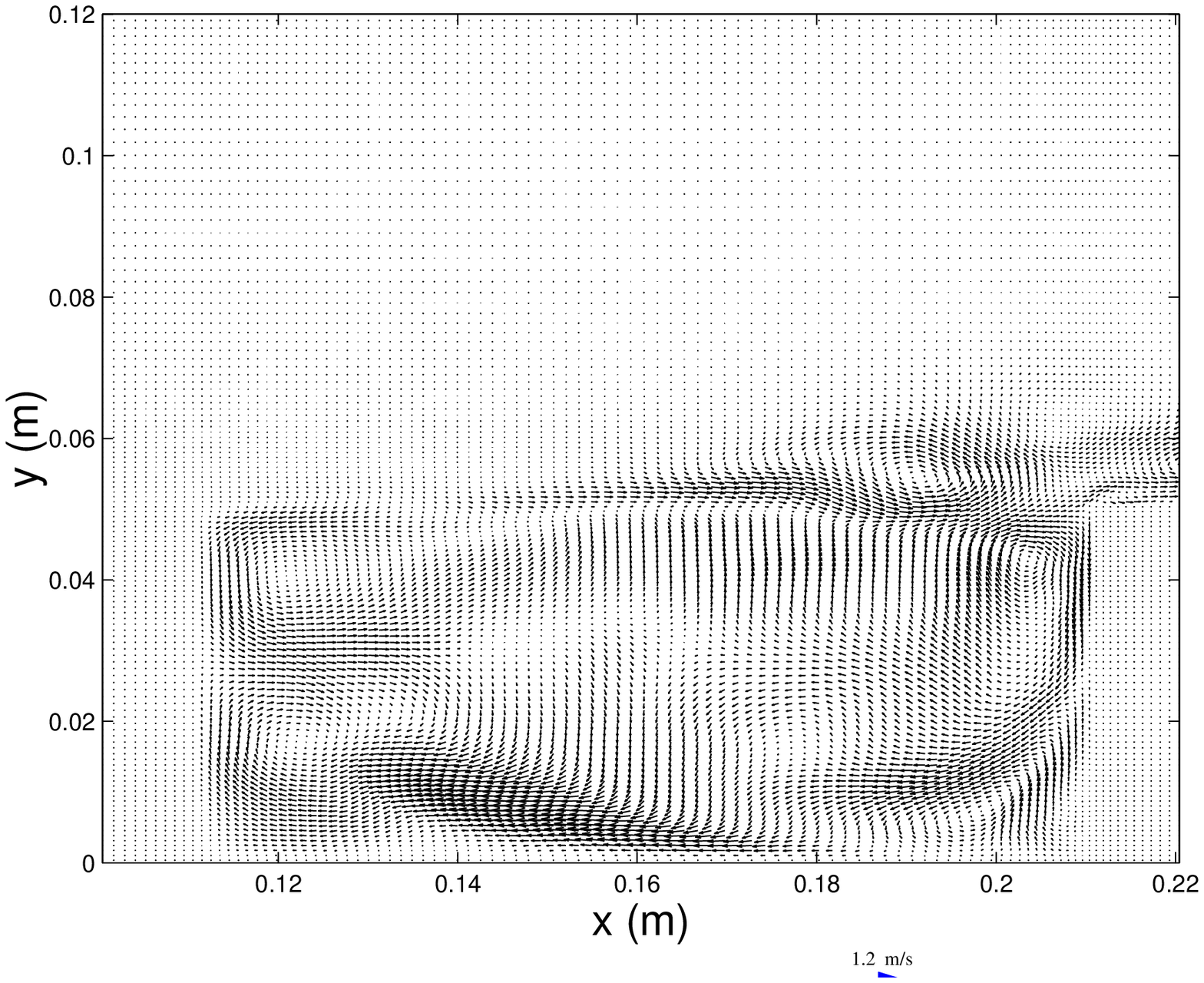}
\hfill
\includegraphics[width=40mm]{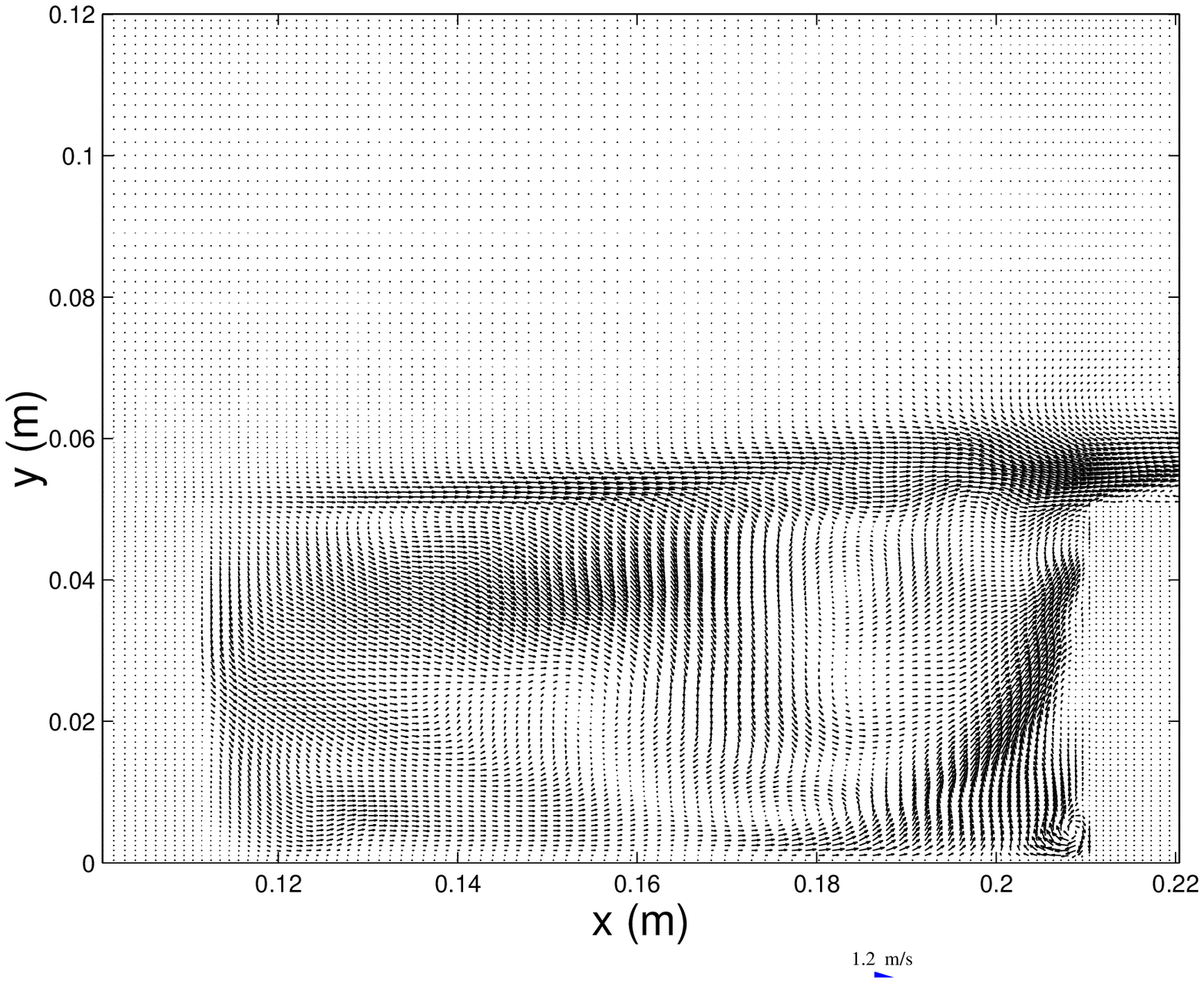}
\caption{Six first spatial eigenmodes (topos $\psi _n(\vec{r})$
with $n=1$ to 6 from top left to bottom right). Arrows represent
the velocity vector in the plane of the mode (here components $x$
and $y$).}\label{fig:podmodes}
\end{figure}

Now we propose to apply our technique so as to identify the
spatial coherent structures $\vec{\psi}_n(\vec{r})$ of the flow,
and track out their dynamical features from their associated
time-dependent amplitudes $\phi _n(t)$. Note that the snapshots
here must be sampled at least at $2f_0\simeq 30$ Hz if we want to
be time-resolved with respect to $f_0$ (Shannon criterion). This
was actually achieved in the numerical simulations.

\begin{figure}[tb]
\includegraphics[width=40mm]{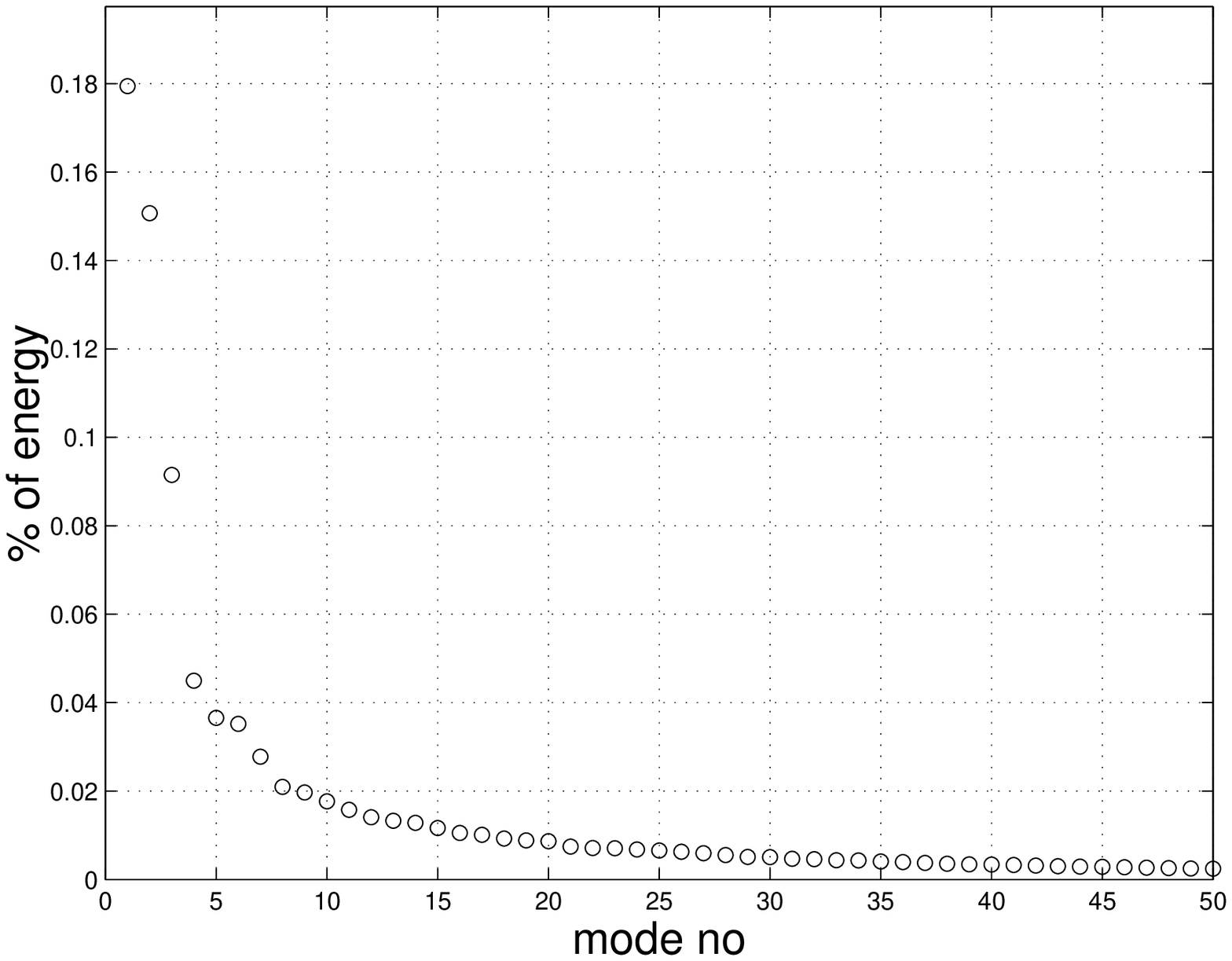}
\hfill
\includegraphics[width=45mm]{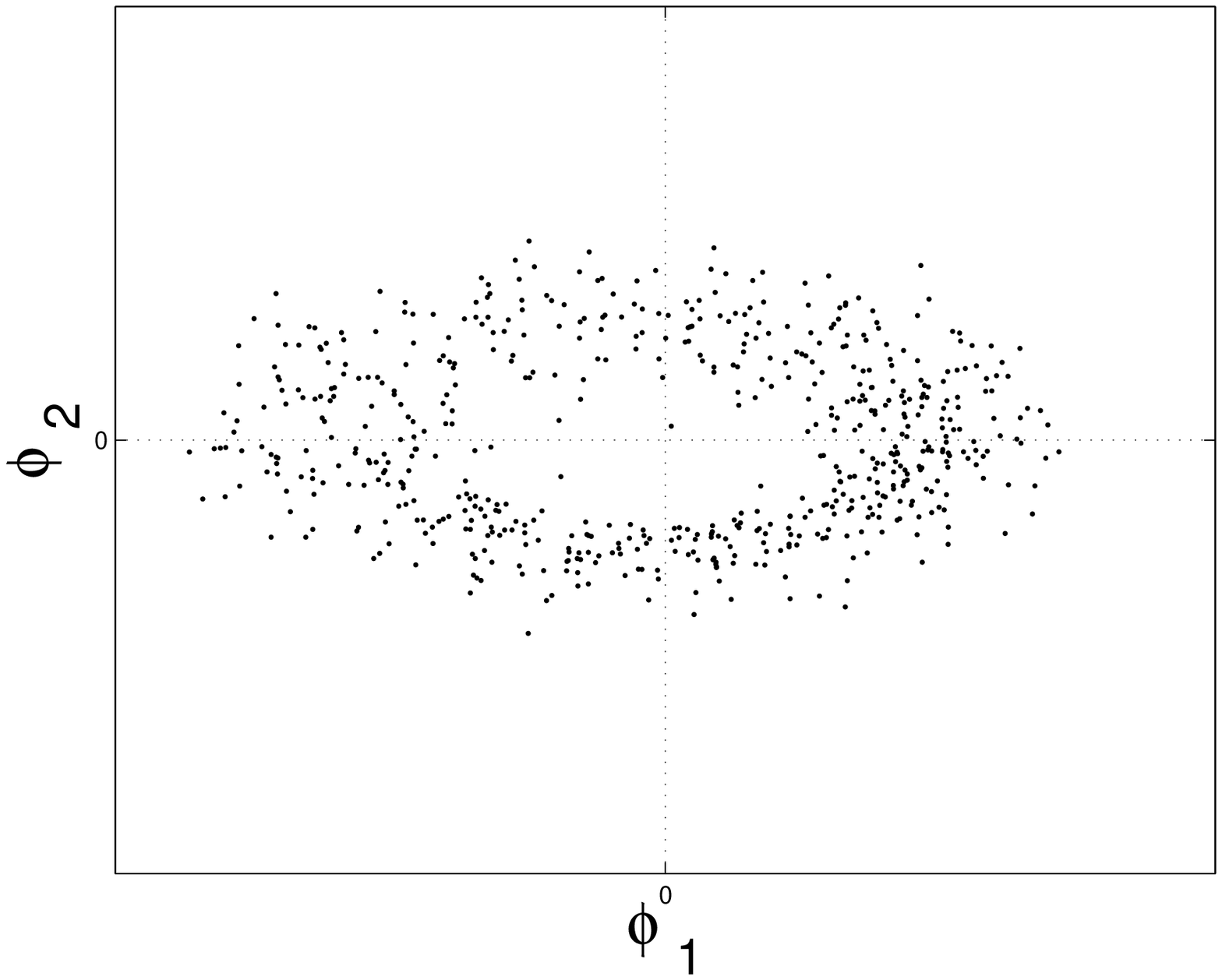}
\caption{Singular value decomposition spectrum. Note the two first
eigenvalues, that are very different (left). Phase portrait of
$\phi_2(t)$ versus $\phi _1(t)$ (right).}\label{fig:podspctrum}
\end{figure}

In Fig.\ref{fig:podmodes} we clearly see that the pod decomposes
the flow into two well-defined areas: one is the mixing layer over
the cavity, essentially captured by the 2 first eigenmodes $\psi
_{1,2}$, the other is the cavity vortices, captured by the higher
order (less energetic) eigenmodes.
The two first modes look very similar, and actually could be phase
squared as expected when the flow is experiencing a global mean
advection (phase squaring resulting in that case from the space
translation invariance) \cite{Lumley}. However, when comparing the
eigenvalues $\lambda _{1,2}$ plotted in Fig.\ref{fig:podspctrum}a,
they appear to be rather different, and not close to each other as
it should be expected in a phase squaring situation. Moreover,
when plotting chronos $\phi _2(t)$ \textit{vs} $\phi _1(t)$
(Fig.\ref{fig:podspctrum}b), a torus is drawn whose dispersion
cannot be explained by numerical noise. Henceforth, it rather
looks like if the two first modes were not two degenerated phase
aspects of a unique ``complex'' mode, but really two different pod
modes, although somehow coupled (so as to produce the torus shape
of Fig.\ref{fig:podspctrum}b). This invoques a symmetry breaking
in the flow advection, most likely due to the downstream corner of
the cavity.
In the discussion on whether the instability is convective or
absolute, note the downstream corner location of the two first
topos $\vec{\psi }_{1,2}$, whose amplitude is vanishing in the
upstream area. This is a strong argument in favor of the
convective nature of the instability, the upstream front of the
instability wavepacket being expected to spread back against the
flow advection in an absolutely unstable situation. A global mode
cannot be completely excluded however \cite{Huerre}.

\begin{figure}[tb]
\includegraphics[width=40mm]{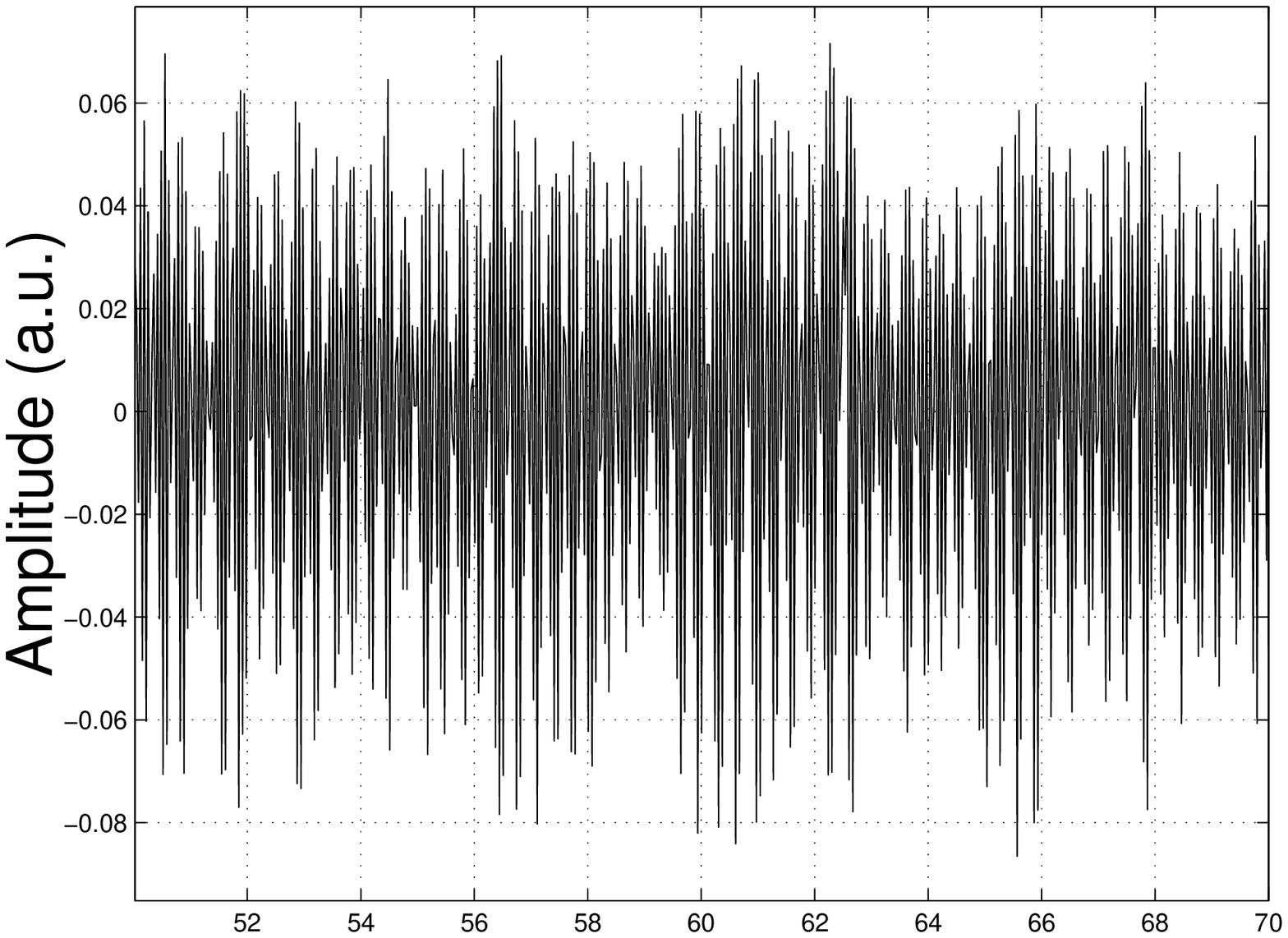}
\hfill
\includegraphics[width=40mm]{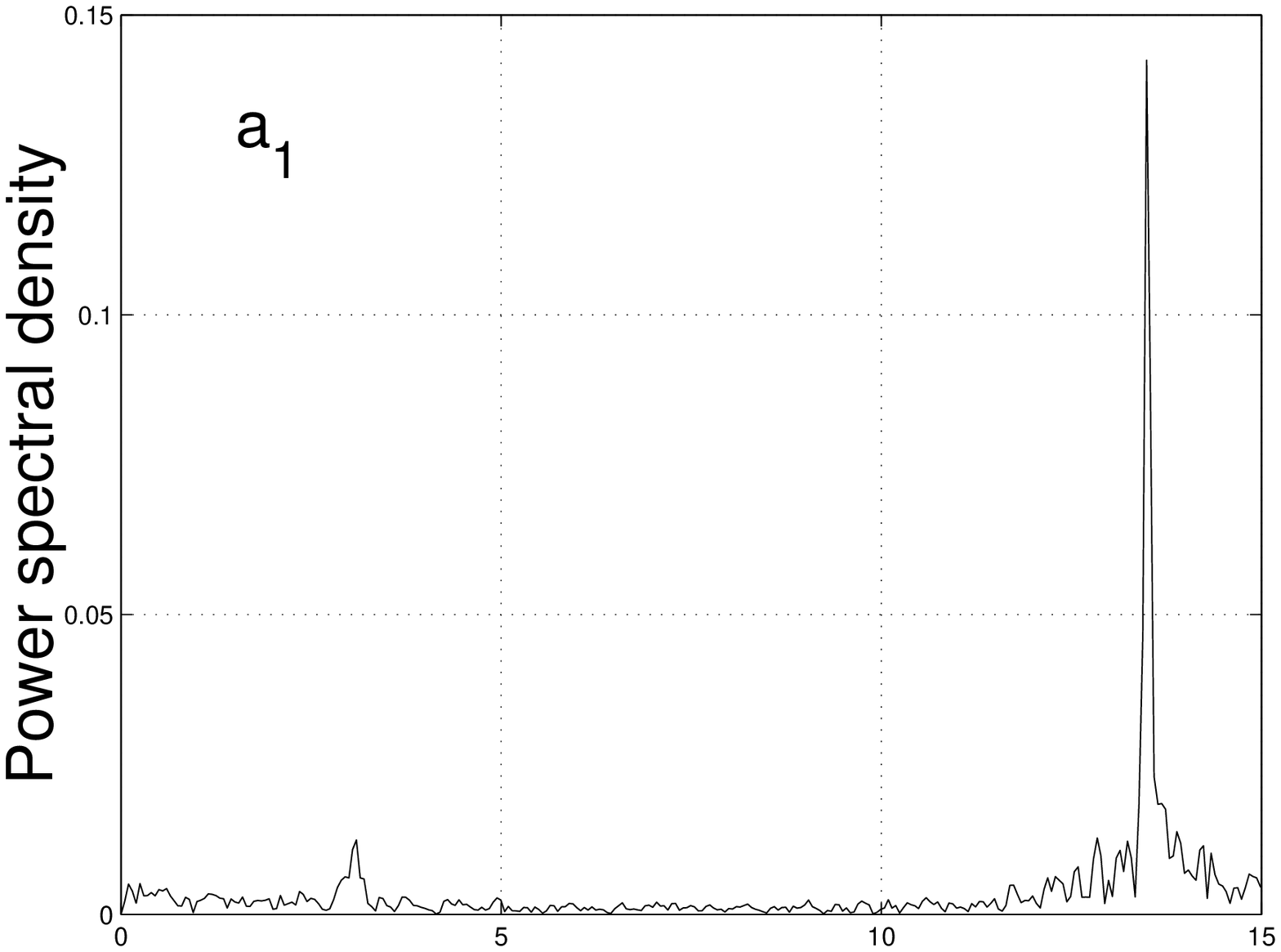}
\newline
\includegraphics[width=40mm]{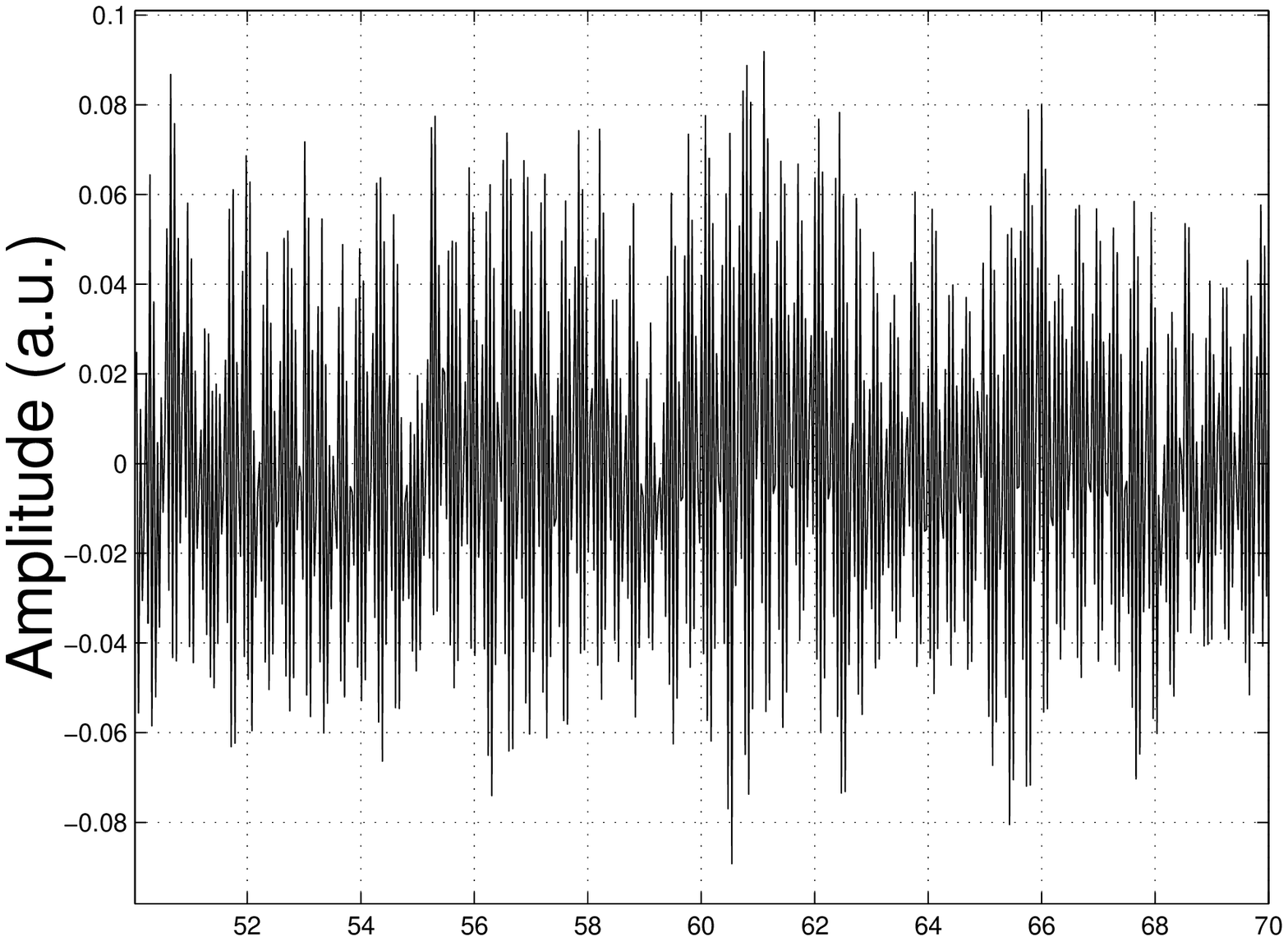}
\hfill
\includegraphics[width=40mm]{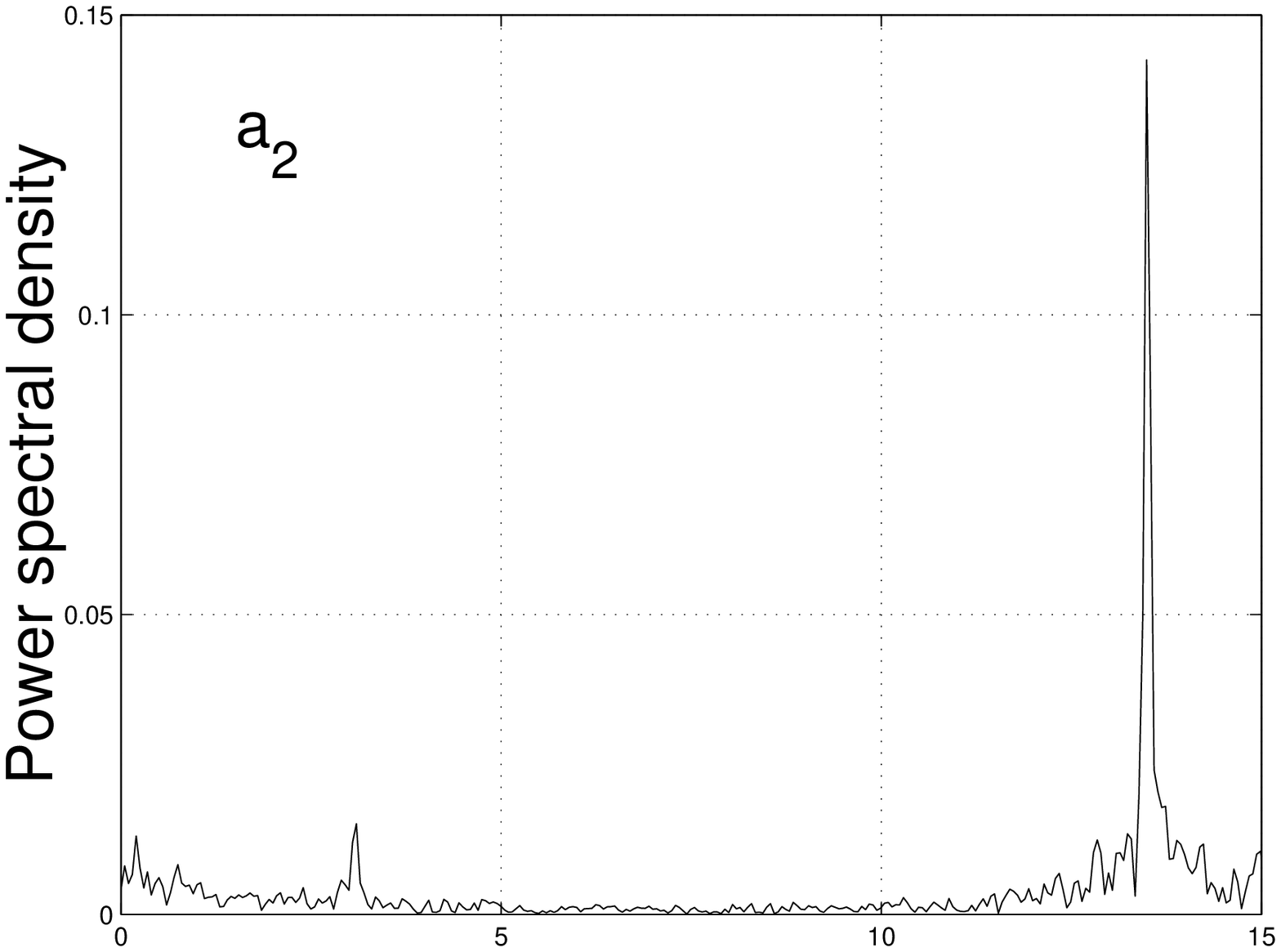}
\newline
\includegraphics[width=40mm]{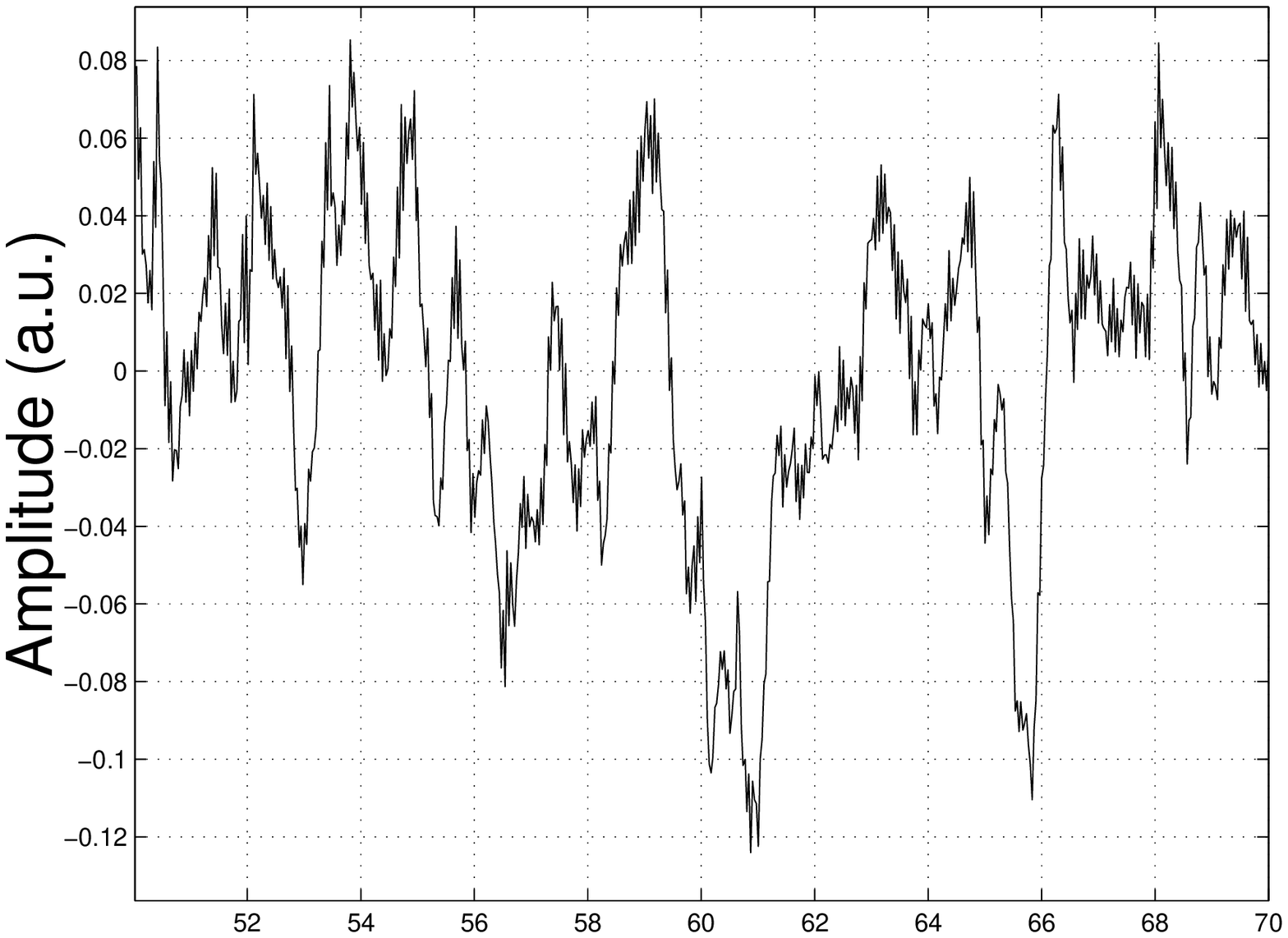}
\hfill
\includegraphics[width=40mm]{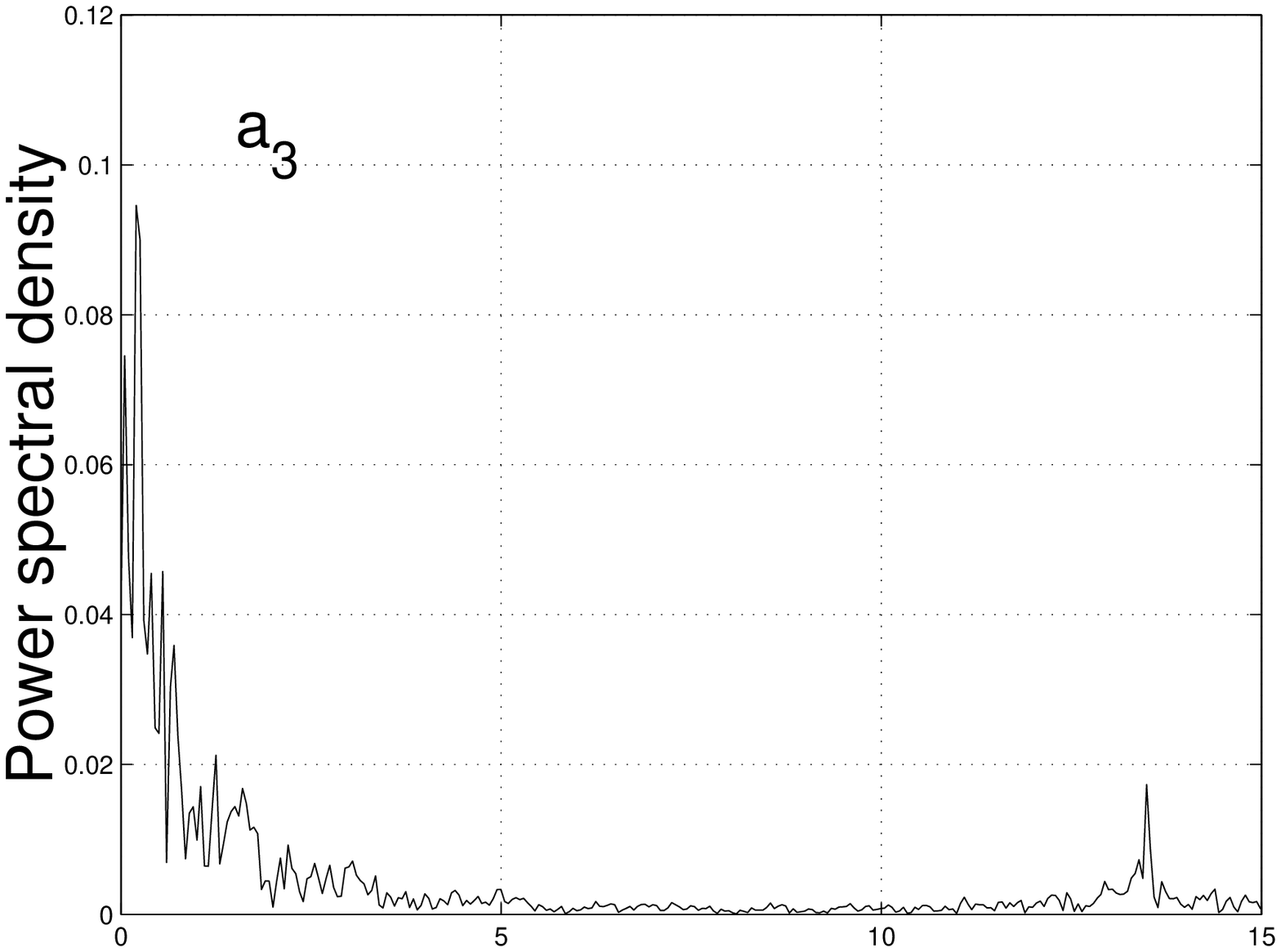}
\newline
\includegraphics[width=40mm]{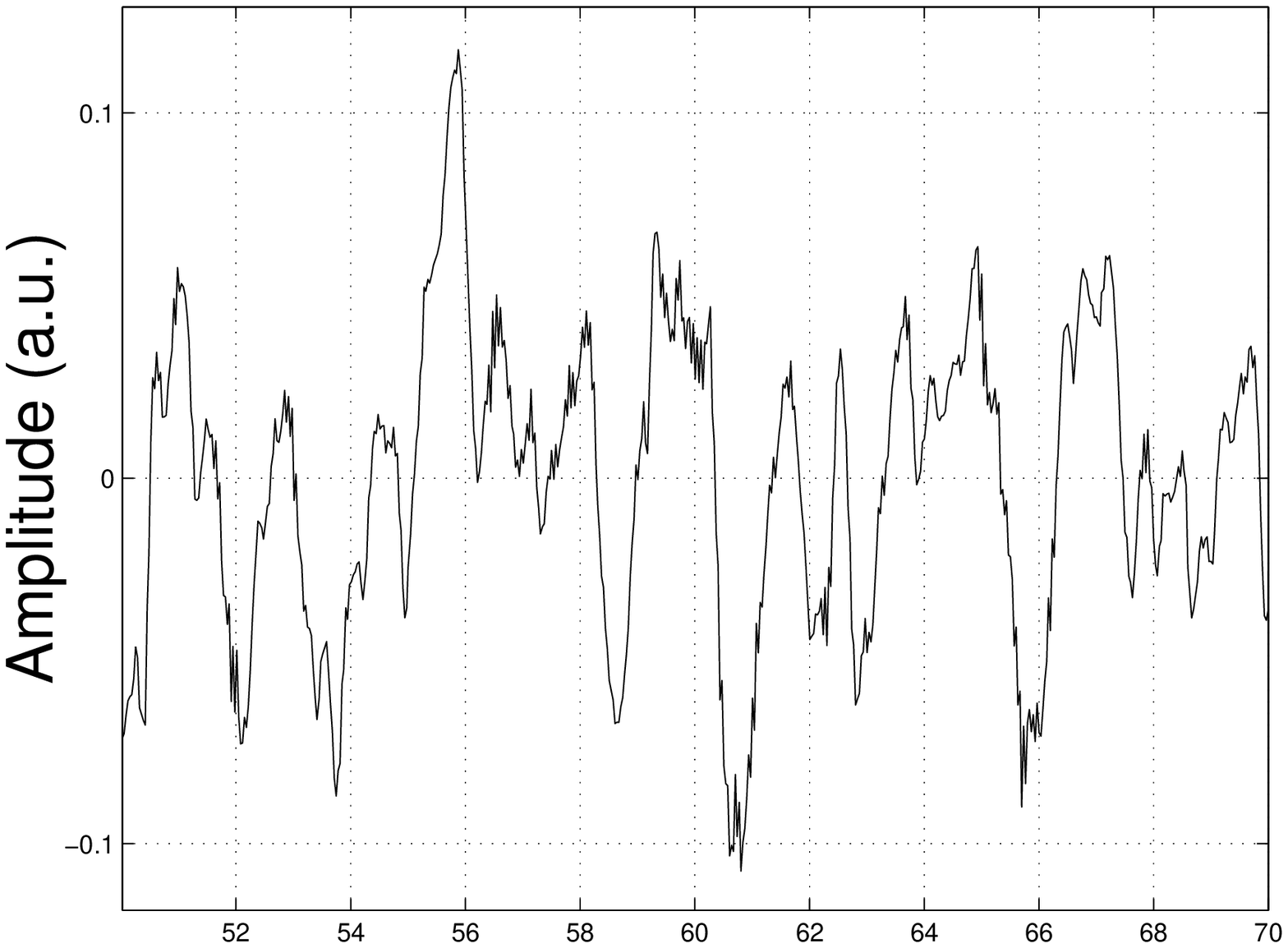}
\hfill
\includegraphics[width=40mm]{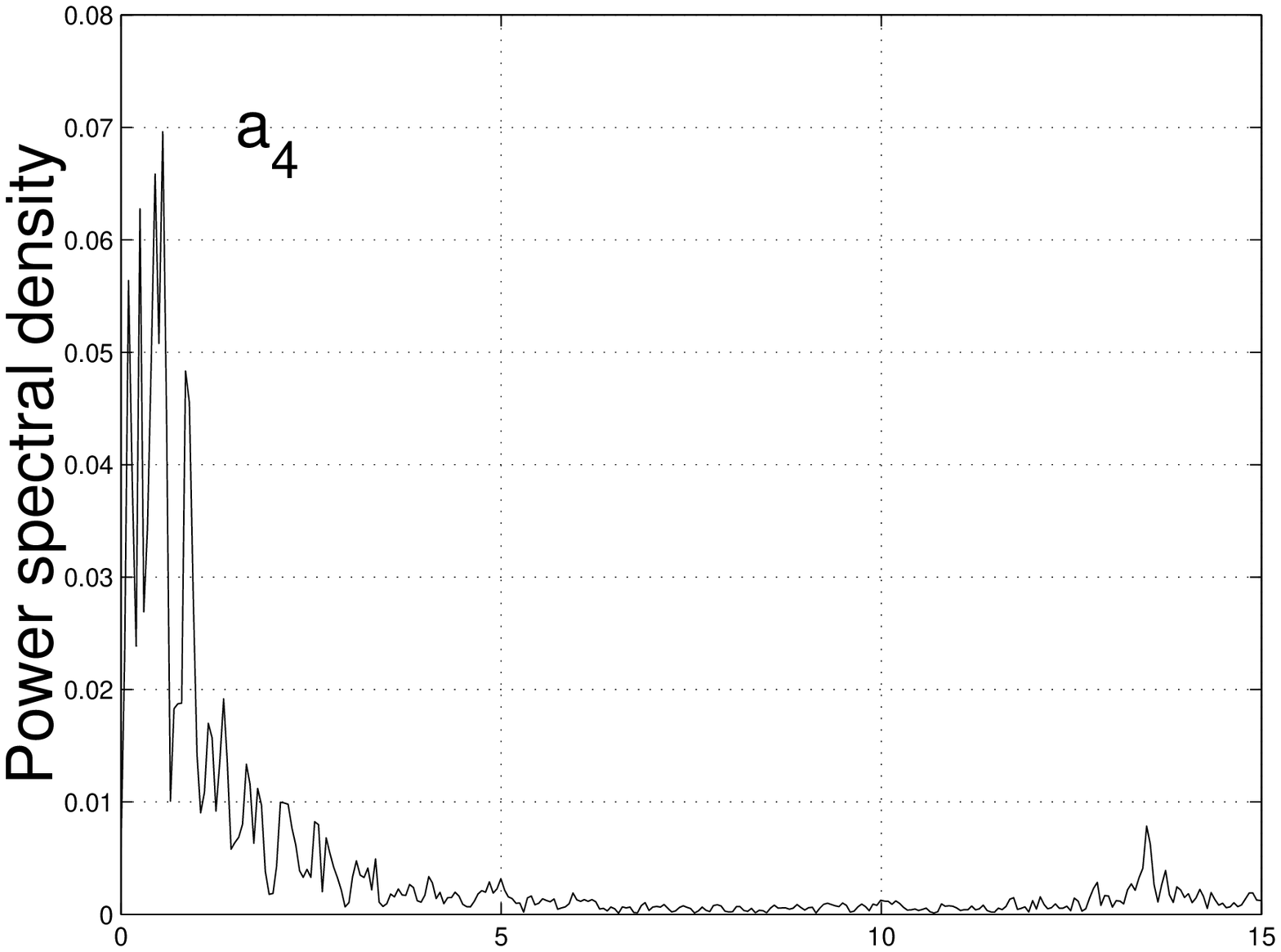}
\newline
\includegraphics[width=40mm]{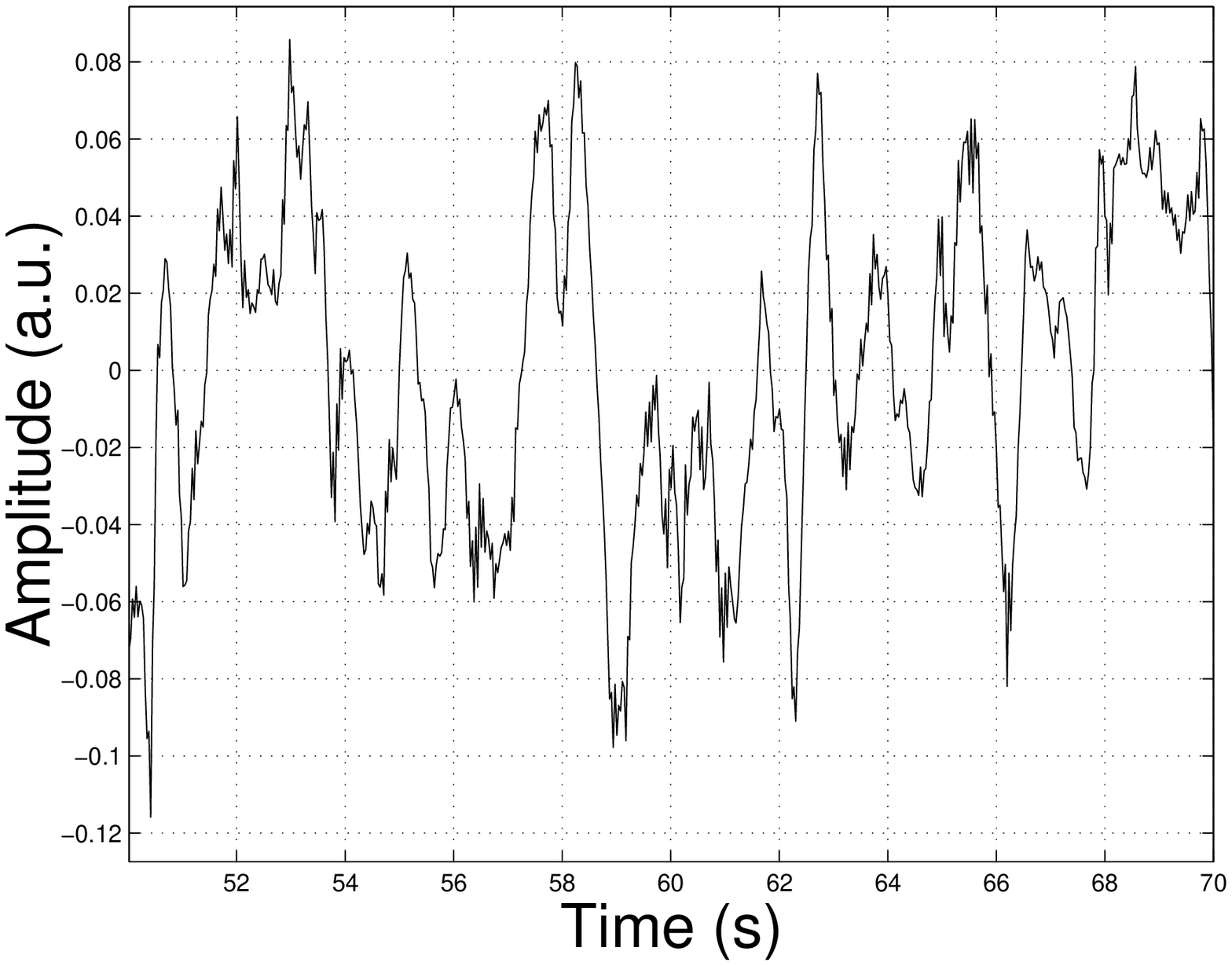}
\hfill
\includegraphics[width=40mm]{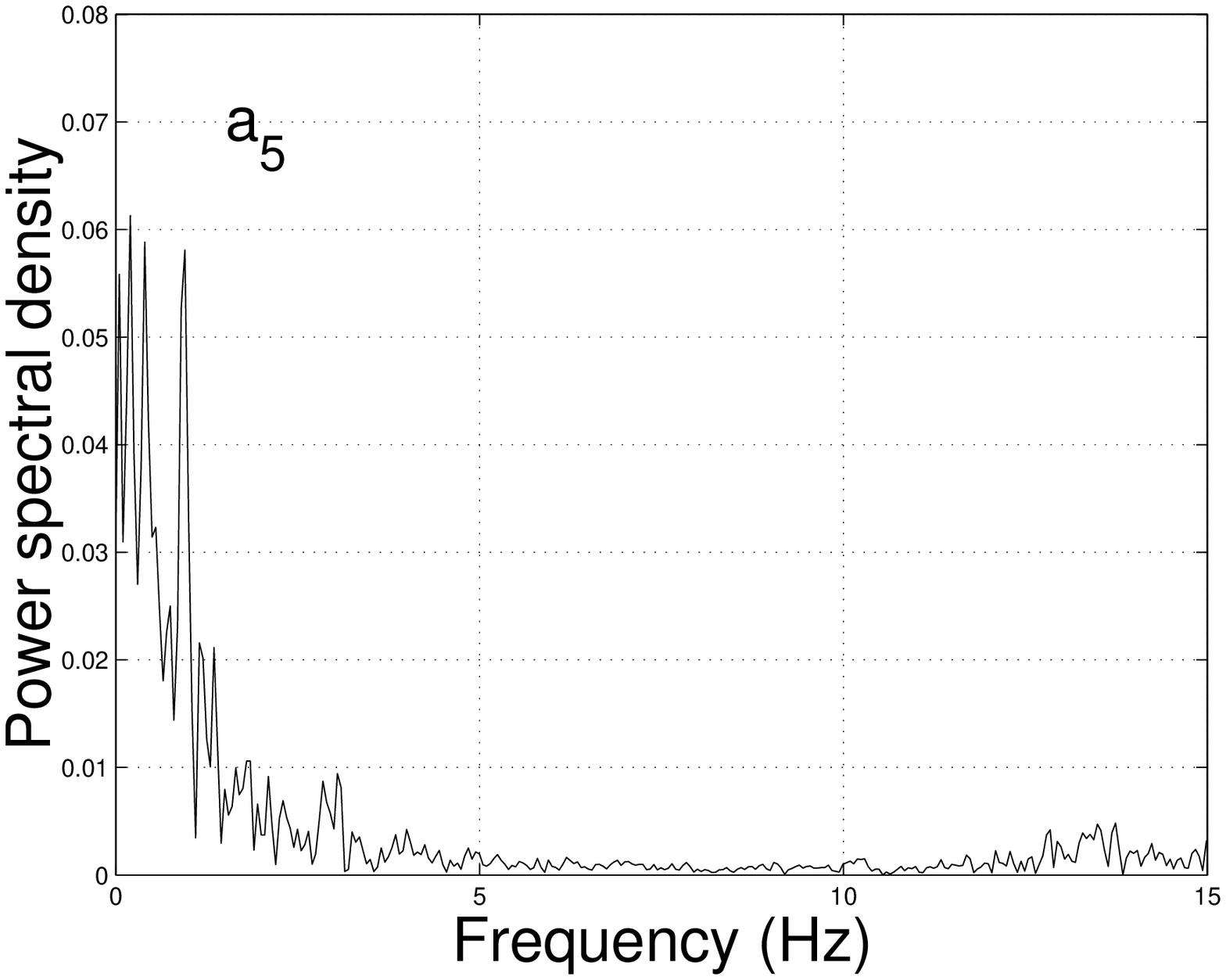}
\caption{Five first time eigenmodes (chronos $\phi _n(t)$ with
$n=1$ to 5 from top to bottom), and their associated power
spectral distribution $a_i$. }\label{fig:chronos}
\end{figure}

In Fig.\ref{fig:chronos} are shown the five first time series
$\phi _n(t)$ and their spectral distribution. We clearly see the
occurrence of the frequency $f_0=13.5$ Hz associated with the two
first chronos, the corresponding topos featuring the coherent
structures contained in the mixing layer. Clearly, the frequency
turns out to be associated with the instability that develops in
the mixing layer. While local time series all produce spectral
components at $f_0$ (see Fig.\ref{fig:tseries}), the pod instead
is able to overcome this global flow coherence and to selectively
associate the spectral components to the adequate spatial coherent
structures. This result henceforth naturally confirms the mixing
layer origin of the most energetic spectral component.

\begin{figure}[tb]
\includegraphics[width=40mm]{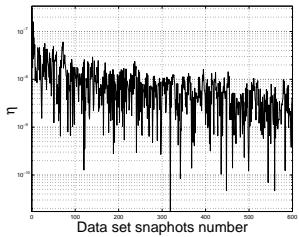}
\caption{Convergence test of mode 1 passing from $p$ samples to
$p+1$ in the data set. See text for description of the $\eta $
criterion definition.}\label{fig:convergence}
\end{figure}

At this step, it might be interesting to briefly discuss some
critical points of the technique. First, because the method is
aimed to track out coherent patterns encountered within a flow
(coherent with respect to the point-wise correlation matrices), it
is important for the statistical flow properties to be stationary.
As a consequence, the data set must possess a sufficiently
important number of independent realizations so as to ensure the
convergence of the decomposition towards the real pod modes. We
have checked that for data set of less than 400 samples, the third
mode fairly mixes both shear layer and cavity structures,
resulting in its time amplitude fourier spectrum to the occurrence
of the $13.5$ Hz-peak --- strongly weakened here in mode 3 when
using 600 samples. Secondly, from an experimental point of view,
each sample composing the data set should share identical
(statistical) properties; as a consequence, when directly working
on instantaneous snapshots of the flow, particule feeding should
remain homogeneous in time, the average intensity and coherent
structure resolution being modified as the feeding is varying ---
therefore biasing the statistical representativity of the samples
\cite{piv}. There are no systematic test to decide whether
statistical convergence has been reached or not. We however
plotted in Fig.\ref{fig:convergence} the average difference $\eta
$ between two modes with respect to the number of snapshots
contained in the data set:
$$\eta (p)=\frac{1}{N}\int _\mathcal{S}\left||\psi _1^{p+1}(\textbf{r})|-|\psi _1^{p}(\textbf{r})|\right|d\textbf{r},$$
where $\psi _n^{p}(\textbf{r})$ is the $n^{th}$-topos computed
using $p$ snapshots in the data set for the single $x$-component
of the velocity. Note that we had to deal with the absolute value
of the topos so as to get rid of the sign, since it was observed
cyclic global sign inversions from $\psi _1^{p}$ to $\psi
_1^{p+1}$, without deep modification of the velocity structure. We
see from Fig.\ref{fig:convergence} that convergence is ensured for
mode 1 with $p\sim 200$ flow realizations.

The study reported here in fact brings another very interesting
insight from an experimental point of view. It indeed shows that,
although the velocity field is spatially fully 3D and
characterized by 3 components \cite{3D}, a 2D pod calculation
(performed in a plane), over 2 velocity components, is able to
separate the two intuitive regions of interest in the flow (namely
the mixing layer and the cavity vortices), which therefore
strongly simplify any experimental protocol, in that a classical
PIV (in a plane, over 2 velocity components) is sufficient to
track out the coherent structures and their dynamical features,
without having to call upon 3D PIV techniques, much heavier. We
have checked that the results were very similar when using 1 or 3
velocity components instead of 2. Moreover, the 3D calculation of
the pod modes confirms all the results provided by the 2D
analyses; 2D cuts out of the 3D modes look very similar to our
(intrinsically) 2D modes, and their amplitude spectral
distribution are comparable as well (see \cite{3D}).

In conclusion, a pod technique has been applied with success to
discriminate the relevant dynamical features of the coherent
structures present in the flow over an open cavity. The processing
time revealed to be of the order of 30 s for about 600 samples of
size 37300 pixels, and grew up to 11 min when applied to about 300
experimental PIV samples of size 241.800 pixels ($N=260\times
465$). However, in most experimental applications, the whole field
resolution, or the whole picture area, are not required to get the
expected results, and it is expected that the technique could
efficiently be applied to a panel of other open flows presenting
self-sustained oscillations.

\texttt{Matlab} programs can be obtained from the authors upon
request.


\begin{thebibliography}{99}
\bibitem{piv}
Note that when dealing with PIV samples, the velocity field is
always defined at each pixel of the picture, whatever the feeding
is (this one just being important for the accuracy in the velocity
estimation), so that most of the requirements may be fulfilled by
the data samples.
\end{thebibliography}
\end{document}